\documentclass{iopart}

\pdfoutput=1

\usepackage{color}
\usepackage{graphicx,epsfig}
\usepackage{bm}
\usepackage{amssymb}

\def\be{\begin{equation}}
\def\ee{\end{equation}}

\def\bea{\begin{eqnarray}}
\def\eea{\end{eqnarray}}

\def\s{\sigma}
\def\th{\theta}
\def\l{\lambda}
\def\sgn{{\rm sgn}}
\def\fr#1{(\ref{#1})}
\def\nn{\nonumber\\}

\begin{document}
\title[Universal corrections to scaling for block entanglement in
spin-$\frac{1}{2}$ XX chains] 
{Universal corrections to scaling for block entanglement in 
spin-$\frac{1}{2}$ XX chains}
\author{Pasquale Calabrese$^1$ and Fabian H.L. Essler$^2$}
\address{$^1$Dipartimento di Fisica dell'Universit\`a di Pisa 
and INFN, Pisa, Italy  

$^2$The Rudolf Peierls Centre for Theoretical Physics, Oxford
University, Oxford OX1 3NP, UK}

\date{\today}

\begin{abstract}
We consider the R\'enyi entropies $S_n(\ell)$ in the one dimensional
spin-1/2 Heisenberg XX chain in a magnetic field. The case $n=1$
corresponds to the von Neumann ``entanglement'' entropy. Using a
combination of methods based on the generalized Fisher-Hartwig
conjecture and a recurrence relation connected to the Painlev\'e VI
differential equation we obtain the asymptotic behaviour, accurate 
to order ${\cal O}(\ell^{-3})$, of the R\'enyi entropies $S_n(\ell)$ for large
block lengths $\ell$. For $n=1,2,3,10$ this constitutes the $3,6,10,48$
leading terms respectively. The $o(1)$ contributions are found to
exhibit a rich structure of oscillatory behaviour, which we analyze in
some detail both for finite $n$ and in the limit $n\to\infty$. 
\end{abstract}

\maketitle
\section{Introduction}
Let $|\Psi\rangle$ be the ground state of an extended quantum mechanical
system and $\rho=|\Psi\rangle\langle\Psi|$ its density matrix.
In order to quantify the bipartite  entanglement in the ground state one divides
the Hilbert space into a part ${\cal A}$ and its complement 
${\cal B}$ and considers the reduced density matrix $\rho_{\cal A}={\rm
  Tr}_{\cal B}\,\rho$ of subsystem ${\cal A}$. 
A measure of the quantum entanglement in the ground-state is provided by
the R\'enyi entropies \cite{Renyi}  
\be
S_n=\frac1{1-n}\ln{\rm Tr}\,\rho_{\cal A}^n\, .
\label{Sndef}
\ee
The particular case $n=1$ of (\ref{Sndef}) is known as the von Neumann
entropy $S_1$ and it is usually called simply {\it entanglement entropy}. 
However, the knowledge of $S_n$ for different $n$ characterizes the
full spectrum of non-zero eigenvalues of $\rho_{\cal A}$ (see e.g.
\cite{cl-08}) and provides significantly more information on the
entanglement than the more widely studied von Neumann entropy. 

Of particular interest is the universal scaling behaviour exhibited by
$S_n$ at quantum critical points. For a one-dimensional critical
system whose scaling limit is described by a conformal field theory
(CFT) of central charge $c$ and ${\cal A}$ being an interval of length $\ell$
embedded in an infinite system, the asymptotic large-$\ell$ behaviour
of the R\'enyi entropies is given by \cite{Holzhey,cc-04,cc-rev} 
\begin{equation}
\label{Renyi:asymp}
S_n(\ell)
\simeq \frac{c}6\left(1+\frac1n \right)\ln \ell+c'_n\,.
\end{equation}
Here $c_n'$ is a non-universal constant. The scaling behaviour
(\ref{Renyi:asymp}) has been verified both analytically and
numerically for a variety of quantum spin chains whose scaling limits
are described by CFTs, see e.g.  
\cite{Vidal,p-04,jk-04,ijk-05,zbfs-06,dmcf-06,ij-08,ncc-08,cg-08,afc-09,gl-rev,gt-10} as well as
in direct field theory calculations \cite{ft}.
In one dimensional systems these entanglement entropies provide
a very useful way for determining the central charge $c$ that
characterizes the behaviour at conformally invariant critical points.
While other methods for determining $c$ such as the finite-size
scaling of the ground state energy 
\cite{cardy,affleck} require the knowledge of certain
non-universal properties such as the velocity of sound, the
large-$\ell$ behaviour of the entanglement provides a \emph{direct}
measure of $c$ as is apparent from Eqn (\ref{Renyi:asymp}). For this
reason a scaling analysis of $S_n$ is increasingly used in numerical
studies of quantum phase transitions in one dimensional systems
\cite{QPT1D,x-10,fibo,song,cv-10,lstn-07,t-08,sun-08,rcla-09,cs-10,hgkm-10,vba-10,ef-10}. 

In such applications $S_n(\ell)$ is computed numerically and the
large-$\ell$ behaviour is then fitted to the form (\ref{Renyi:asymp}).
It has been observed that the asymptotic result is sometimes obscured
by large, and often oscillatory, corrections to scaling
\cite{lsca-06,ccen-10,song,cv-10}. In Ref. \cite{ccen-10}, on the
basis of both exact and numerical results, it has been argued that
these corrections are in fact \emph{universal} and encode information
about the underlying CFT beyond what is captured by the central
charge alone. More precisely, they give access to the scaling dimensions
of some of the most relevant operators in the underlying CFT.  
This conjecture of Ref. \cite{ccen-10} has been recently confirmed
by using perturbed CFT arguments \cite{cc-10}. 

A precise characterization of the subleading terms in $S_n(\ell)$ is
then desirable for two reasons. First, the knowledge
of their structure will be helpful when using (\ref{Renyi:asymp}) to
extract the central charge from numerical computations of $S_n(\ell)$.
Second, the subleading terms can be used to infer the scaling
dimensions of certain operators in the CFT characterising the
quantum critical point.
This motivates the present study, in which we significantly extend our
recent calculation \cite{ccen-10} of the subleading corrections to the
R\'enyi entropies in the XX chain. 
\subsection{Spin-1/2 XX Chain}
The Hamiltonian of the XX model on an infinite one dimensional chain is
\be
H = -  \sum_{l=-\infty}^\infty {1\over 2} [  \sigma^x_l \sigma^x_{l+1} +
  \sigma^y_l \sigma^y_{l+1}] - h \sigma^z_l , 
\label{HXX}
\ee
where $\sigma_l^{x,y,z}$ are the Pauli matrices at site $l$. The
Jordan-Wigner transformation 
\be
c_l=\left(\prod_{m<l} \s^z_m\right)\frac{\s^x_l+i\s_l^y}2\,,
\ee
maps this model to a quadratic Hamiltonian of spinless fermions 
\be
H =
-\sum_{l=-\infty}^\infty  c_l^\dagger c_{l+1}  + c_{l+1}^\dagger
c_{l} + 2h \left(c_l^\dagger c_{l}-\frac{1}{2}\right). 
\label{Hfermi}
\ee
Here $h$ represents the chemical potential for the spinless fermions
$c_l$, which satisfy canonical anti-commutation relations
$\{c_l,c^\dagger_m\}=\delta_{l,m}$. The Hamiltonian (\ref{Hfermi}) is
diagonal in momentum space and for $|h|<1$ the ground-state is a
partially filled Fermi sea with Fermi-momentum 
\be
k_F=\arccos |h|.
\ee
In the following we will always assume that $|h|<1$ so that we are
dealing with a gapless theory. 
\subsection{Entanglement Entropy of the XX chain: Jin-Korepin Result}
A key result regarding on entanglement measures in the XX chain is due
to Jin and Korepin \cite{jk-04}, who obtained the leading large-$\ell$
behaviour of $S_n$. Their result takes the form
\be
S_n^{JK}(\ell)=\frac16\left( 1+\frac1n\right)\ln (2\ell |\sin
k_F|)+E_n\,, 
\label{SnJK}
\ee
where the constant $E_n$ has the integral representation
\be\fl
E_n=\left(1+\frac1n\right)\int_0^\infty \frac{dt}t 
\left[\frac{1}{1-n^{-2}}
\left(\frac1{n\sinh t/n}-\frac1{\sinh t}\right)
\frac1{\sinh t}-\frac{e^{-2t}}6\right]\, .
\label{cnp}
\ee
The objective of our work is to determine the subleading corrections
to $S_n^{JK}(\ell)$ for large, finite block lengths $\ell$. It is
therefore convenient to introduce quantities $d_n(\ell)$ 
\be
 d_n(\ell)\equiv S_n(\ell)-S_n^{JK}(\ell)\,,
\ee
to which we will refer throughout.
\vskip .25cm
The remainder of this paper is organized as follows. 
For the sake of clarity we first present a summary of our results in
section \ref{sec:summary}. When then turn to the details of our
derivations. In section \ref{defsec} we briefly review one of our key
tools, the \emph{generalized Fisher-Hartwig conjecture}. The latter is
used in section \ref{fhsec} to determine all ``harmonic'' corrections
to the R\'enyi entanglement entropies.
In order to go beyond the generalized Fisher Hartwig conjecture we
utilize recent developments related to Random Matrix Theory. These are
introduced in section \ref{rmsec} and used to determine
``non-harmonic'' terms in  the asymptotic expansion for 
the von Neumann and R\'enyi entropies in sections  \ref{vnsec} and
\ref{Rnsec} respectively. 
Comparisons between our analytic expansion and numerical results are
presented in section \ref{numsec}. 

\section{Summary of Results}
\label{sec:summary}
This section contains a summary of our results.
\subsection{R\'enyi Entropies of the XX chain: General Result}
Our full result for $d_n(\ell)$ can be cast in the form
\bea\fl
d_n(\ell)&=&\frac{2}{n-1}{\sum_{p,q=1}^\infty}
(-1)^{p} L_k^{-\frac{2p(2q-1)}{n}}\big(Q_{n,q}\big)^p \left[
\frac{\cos(2 k_F \ell p)}{p}
+  \frac{A_q  \sin(2k_Fp\ell)}{L_k}
\right.  \nn \fl&&\qquad\qquad\qquad\qquad\qquad\qquad \qquad\left.
+\frac{[B^{(n)}_{p,q}e^{2ipk_F\ell}+{\rm h.c.}]}{L_k^2}
\right]
\nn\fl&&
+\frac{1}{L_k^2}\frac{n+1}{285
  n^3}\left(15(3n^2-7)+(49-n^2)\sin^2k_F\right) +{\cal
  O}\Big(L_k^{-3}\Big)\,,
\label{fullresult} 
\eea
where 
\bea
L_k&=&2\ell |\sin k_F|\,,
\label{Lk}\\
A_q&=&\left[1+3\left(\frac{2q-1}{n}\right)^2\right]\cos k_F \,,\\
Q_{n,q}&=&\left[\frac{\Gamma(\frac{1}{2}+\frac{2q-1}{2n})}
{\Gamma(\frac12-\frac{2q-1}{2n})}\right]^2\,,
\label{Qn}
\eea
\bea
B^{(n)}_{p,q}&=&\frac{2q-1}{6n}\left[\big(5+7\frac{(2q-1)^2}{n^2}\big)
\sin^2(k_F)-15\big(\frac{(2q-1)^2}{n^2}+1\big)\right]\nn
&&
-\frac{p}{4}\left[\big(1+3\frac{(2q-1)^2}{n^2}\big)
\cos(k_F)\right]^2.
\label{Bpq}
\eea

The leading contribution to $d_n(\ell)$ has already been announced
in Ref. \cite{ccen-10} and is given by
\be
d_n(\ell)=
\frac{ 2 \cos(2 k_F \ell) }{1-n} (2\ell |\sin k_F |)^{-2/n}
Q_{n,1}
+{\cal O}\big(\ell^{-\min[4/n,2]}\big).
\label{Sncorrintro}
\ee
\subsection{R\'enyi Entropies of the XX chain: explicit results for 
$S_2(\ell)$ and $S_3(\ell)$}
In the special cases $n=2$ and $n=3$ our results read
\bea\fl
d_2(\ell)&=&-\frac{2 Q_{2,1}\cos(2 k_F \ell)}{L_k}+
\frac1{L_k^{2}}\bigg[
Q_{2,1}^2\cos(4k_F\ell)\nn \fl&&
\qquad\qquad- \frac{7 Q_{2,1} \cos k_F}2 \sin (2k_F \ell)
+ \frac{5+3\sin^2 k_F}{64} \bigg]+{\cal O}\Big(L_k^{-3}\Big)
\label{S2} ,
\eea 
and 
\bea
\fl
d_3(\ell)&=&-\frac{ Q_{3,1}\cos(2 k_F \ell)}{L_k^{2/3}}
+\frac{Q_{3,1}^2\cos(4k_F\ell)}{2L_k^{4/3}}
-\frac{4 Q_{3,1} \cos k_F}{3 L_k^{5/3}} \sin (2k_F \ell)\nn\fl
&&
-\frac{Q_{3,1}^3\cos(6k_F\ell)}{3L_k^{2}}
+2\frac{15+2\sin^2k_F}{243 L_k^2}
+\frac{4}{3}\frac{\cos(k_F)Q_{3,1}^2\sin(4k_F\ell)}{L_k^{7/3}}\nn\fl
&&
+\frac{Q_{3,1}^4 \cos(8k_F \ell)}{4L_k^{8/3}}
+\frac{2Q_{3,1}(111-62\sin^2(k_F))\cos(2k_F\ell)}{81L_k^{8/3}}
+{\cal O}\Big(L_k^{-3}\Big)\,.
\label{S3}
\eea

\subsection{R\'enyi Entropies of the XX chain: limit $n\to\infty$}

In the limit $n\to\infty$ infinitely many terms in (\ref{fullresult})
combine to generate a logarithmic contribution, whose general
expression is is given in Eq. (\ref{logcorr}). It assumes a
particularly simple form at half-filling $k_F=\pi/2$
\be
d_\infty(\ell)\simeq
\frac{\pi^2}{24\ln(2b \ell)}
\cases{\displaystyle
2 & $\ell$ {\rm odd} \,,\\ \displaystyle
-1 & $\ell$ {\rm even}\,,
}
\ee
where $b=\exp(-\Psi(1/2))\approx 7.12429$. 

\subsection{von Neumann Entropy of the XX chain}
In the special case $n=1$ corresponding to the von Neumann entropy
all oscillating contributions to (\ref{fullresult}) vanish.
This explains why it is easier to determine
the central charge from $S_1$ than from R\'enyi entropies with $n\geq
2$ (this is no longer true in the presence of boundaries
\cite{lsca-06}, where it is found that oscillations persist in the
limit $n\to1$). Specializing Eq. (\ref{fullresult}) to $n=1$ we obtain 
\be
S_1
\simeq \frac{1}3\ln \ell+c'_1-\frac1{12\ell^2}\left(\frac15+\cot k_F^2\right)\,.
\label{vNintro}
\ee
In this expression, the $\ell^{-2}$ power-law behaviour is \emph{universal}
\cite{ccen-10,ncc-08}.

\section{Entanglement entropy in the XX model}
\label{defsec}
Let us return to the spin-1/2 XX model on an infinitely long chain
(\ref{HXX}). The reduced density matrix of a block of $\ell$
contiguous sites can be expressed as
\be
\rho_{\cal A}=\det{C}\
\exp\left(\sum_{j,l\in{\cal A}}\big[\ln(C^{-1}-1)\big]_{jl}
c^\dagger_jc_l\right),
\ee
where the \emph{correlation matrix} $C$ has matrix elements
\be
C_{nm}= \langle c_m^\dagger c_n \rangle=  \frac{\sin k_F (m-n)}{\pi(m-n)}.
\label{Cnm}
\ee
As a real symmeetric matrix $C$ can be diagonalized by a unitary
transformation
\be
U C U^\dagger\equiv \delta_{lm} (1+\nu_m)/2. 
\ee
This implies that the reduced density matrix $\rho_\ell$ is
uncorrelated in the transformed basis. The R\'enyi entropies can
be expressed in terms of the eigenvalues $\nu_l$ as 
\be\fl
S_n(\ell)=\sum_{l=1}^\ell e_n(\nu_l)\,, \quad {\rm with }\quad
e_n(x)=\frac1{1-n}\ln \left[\left(\frac{1+x}2\right)^n+ \left(\frac{1-x}2\right)^n\right]\,.
\label{Sn}
\ee
More details about this procedure can be found in e.g.
Refs. \cite{Vidal,gl-rev,pe-rev}. We note that the above construction
refers to the block entanglement of fermionic degrees of
freedom. However, in the case considered here, the non-locality
induced by the Jordan-Wigner transformation does not affect the
reduced density matrix. In fact, it can be seen to mix only spins
inside the block. This ceases to be the case when two or more disjoint
intervals are considered \cite{atc-09,ip-09} and other techniques
need to be employed \cite{fc-10} in order to recover CFT predictions
\cite{fps-09,cct-09,atc-09}.  

The representation (\ref{Sn}) is particularly convenient for numerical
computations: the eigenvalues $\nu_m$ of the $\ell\times\ell$
correlation matrix $C$ are determined by standard linear algebra
methods and $S_n(\ell)$ is then computed using Eq. (\ref{Sn}). 
In order to obtain the universal behaviour in the limit of large block
lengths $\ell\to\infty$ we follow Ref. \cite{jk-04}. We introduce the
determinant 
\be
D_\ell(\l)=\det\big((\l+1) I -2 C\big)\equiv \det(G)\,.
\ee
In the eigenbasis of $C$ the determinant is simply a polynomial of
degree $\ell$ in $\l$ with zeroes $\{\nu_j|j=1,\ldots,\ell\}$, i.e.
\be
D_\ell(\l)=\prod_{j=1}^\ell (\l-\nu_j).
\ee
This implies that the R\'enyi entropies have the integral representation
\be
S_n(\ell)= \frac1{2\pi i}\oint d\l\ e_n(\l) \frac{d\ln
  D_\ell(\l)}{d\l}\ ,
\label{Snint}
\ee
where the contour of integration encircles the segment $[-1,1]$.
The matrix $G$ is a $\ell\times \ell$ Toeplitz matrix, i.e. its
matrix elements depend only on the difference between row and column
indices 
\be
G_{jk}=g_{j-k}\ .
\ee
In the theory of Toeplitz matrices an important role is played by the
Fourier transform $g(\theta)$ of $g_l$ 
\bea
g_l&=&\int_0^{2\pi}\frac{d\theta}{2\pi}\ e^{il\theta}\ g(\theta).
\eea
The function $g(\theta)$ is called \emph{symbol} and in our case takes
the form
\be
g(\theta)=
\cases{
\lambda+1 & $\theta\in[k_F,2\pi-k_F]$\\
\lambda -1 & $\theta\in [0,k_F]\cup[2\pi-k_F,2\pi]$\ .
}
\label{symbol0}
\ee
On the interval $[0,2\pi]$ the function $g(\th)$ has two
discontinuities at $\th_1=k_F$ and $\th_2=2\pi-k_F$. 
\subsection{The generalized Fisher-Hartwig conjecture}
The Fisher-Hartwig conjecture \cite{fh-c} gives
the asymptotic behaviour of the determinant of a Toeplitz matrix in
the limit where the dimension $\ell$ of the matrix becomes large.
This has been used by Jin and Korepin \cite{jk-04} to derive
the leading large $\ell$ asymptotic behaviour of the R\'enyi
entanglement entropies. As stressed in Ref. \cite{jk-04}, for the
Toeplitz matrices defined by the symbol (\ref{symbol0}) the
Fisher-Hartwig conjecture has been proven by Basor \cite{bm-94}. 

In order to employ the Fisher-Hartwig conjecture one needs to express
the symbol $g(\theta)$ of a Toeplitz matrix in the form
\be
g(\th)=f(\th)\prod_{r=1}^R e^{ib_r[\th-\th_r-\pi\sgn(\th-\th_r)]}
\left(2-2\cos(\th-\th_r)\right)^{a_r},
\label{symbol}
\ee
where $R$ is an integer, $a_r$, $b_r$ and $\theta_r$ are constants and
$f(\theta)$ is a smooth function with winding number zero.
The Fisher-Hartwig conjecture then states that the large-$\ell$
asymptotic behaviour of the corresponding Toeplitz determinant
is given by 
\be
D_\ell\sim F[f(\theta)]^\ell\left(\prod_{j=1}^R \ell^{a_j^2-b_j^2} \right) E\,,
\ee 
where $F[f(\theta)]=\exp(\frac1{2\pi} \int_0^{2\pi} d\th \ln f(\th))$
and $E$ is a known function of 
$f(\th)$, $a_r$, $b_r$, and $\th_r$. In our case it is straightforward
to express the symbol in the canonical form
(\ref{symbol}). As $g(\theta)$ has two discontinuities in $[0,2\pi)$
we have $R=2$. It is useful to define a function 
\bea
\beta_\lambda&=&\frac{1}{2\pi
  i}\ln\left[\frac{\lambda+1}{\lambda-1}\right],
\eea
where the branch cut of the logarithm is chosen such that
\be
-\pi\leq \arg\left[\frac{\lambda+1}{\lambda-1}\right]< \pi.
\ee
Inserting the ansatz
\be
b_1=-b_2\ ,\quad a_{1,2}=0\ ,\quad f(\th)=f_0={\rm const}
\ee
into \fr{symbol} gives
\be
g(\theta)=
f_0e^{2ib_2k_F} 
\cases{
1& $\theta\in[k_F,2\pi-k_F]$\\
e^{-2\pi ib_2} & $\theta\in [0,k_F]\cup[2\pi-k_F,2\pi]$\ .
}
\label{gth2}
\ee
Comparing (\ref{gth2}) to (\ref{symbol0}) we conclude that we require
\be
b_2=\beta_\lambda+m\ ,
\ee
where $m$ is an arbitrary integer number. We further identify
\be
f_0=(\lambda+1)e^{-2ib_2k_F}=(\lambda+1)e^{-2ik_F m}e^{-2ik_F\beta_\l}.
\ee
The integer $m$ labels the different inequivalent
\emph{representations} of the symbol $g(\th)$, see \cite{bm-94}. 
In their work Jin and Korepin employed the Fisher-Hartwig conjecture
for the $m=0$ representation and obtained the following result for the
large-$\ell$ asymptotics of $D_\ell(\lambda)$ \cite{jk-04}
\be\fl
D_\ell^{JK}(\l)\sim
\left[(\l+1)\left(\frac{\l+1}{\l-1}\right)^{-\frac{k_F}\pi}\right]^\ell \
L_k^{-2\beta^2(\l)} G^2(1+\beta_\l)G^2(1-\beta_\l)\,,
\label{Djk}
\ee
where $L_k= 2 \ell |\sin k_F|$ has been introduced in (\ref{Lk}).
Inserting (\ref{Djk}) into (\ref{Snint}) and carrying out the integral
leads to the result for the asymptotic behaviour of the
R\'enyi entropy reported in Eq. (\ref{SnJK}).
Expression (\ref{SnJK}) provides the leading behaviour of $S_n(\ell)$
for large block lengths $\ell$. It is the purpose of our work to
determine (universal) subleading contributions to (\ref{SnJK}).
This is achieved by noting that for the case when the symbol
$g(\theta)$ has several inequivalent representations labelled by an
integer $m$ the asymptotics of the corresponding Toeplitz determinant
is given by the so-called  {\it generalized Fisher-Hartwig conjecture}
(gFHC) \cite{bm-94}, which reads 
\be
D_\ell(\l)\sim\sum_m e^{l_0^{(m)}\ell}
\ell^{-\sum_{r=1}^2(b^{(m)}_r)^2}E^{(m)}. 
\label{Drep}
\ee
In our case, the various parameters in (\ref{Drep}) are given by
\bea
l_0^{(m)}&=&\ln(f_0^{(m)})=\ln(\lambda+1)-2ik_F\beta_\lambda-2ik_Fm,\\ 
b_2^{(m)}&=&-b_1^{(m)}=\beta_\lambda+m\,,
\eea
where $m$ are integers and
\be\fl
E^{(m)}=\left[2-2\cos(2k_F)\right]^{-(\beta_\lambda+m)^2}
\left[ G(1+\beta_\lambda+m)G(1-\beta_\lambda-m)\right]^2.
\ee
Here $G(z)$ is the Barnes $G$-function \cite{BG}. 
We note that the gFHC has been used to determine the large-distance
asymptotics of various two-point correlation functions in Refs
\cite{ov,fa-05}.
Important properties of the gFHC in our case are
\begin{enumerate}
\item{}
The exponential increase is representation independent and governed by
the exponent
\be
{\rm Re}\big(l_0^{(m)}\big)={\rm Re}\left[\ln(\lambda+1)\right]-\frac{k_F}{\pi}\ 
{\rm Re}\left[\ln\left[\frac{\lambda+1}{\lambda-1}\right]\right]
\ee
\item{} 
The leading oscillatory behaviour depends on the representation and is given by
\be
{\rm Im}\big(l_0^{(m)}\big)={\rm
  Im}\left[\ln(\lambda+1)\right]-\frac{k_F}{\pi}\  
{\rm Im}\left[\ln\left[\frac{\lambda+1}{\lambda-1}\right]\right]
-2k_F m.
\ee
\item{} 
The power law correction depends on the representation and is
characterized by the exponents
\be
\alpha_m=\big(b_1^{(m)}\big)^2+\big(b_2^{(m)}\big)^2
=-2(\beta_\lambda+m)^2.
\ee
The real parts of these exponents are
\be
{\rm Re}(\alpha_m)=-2{\rm Re}(\beta_\l^2)-2m(m+2{\rm Re}(\beta_\l)).
\label{realpha}
\ee
In conjunction with the inequality $-1\leq 2{\rm Re}(\beta_\l))<1$
this establishes that
\be
{\rm Re}(\alpha_m)\leq{\rm Re}(\alpha_0).
\label{alpham}
\ee
Equality in (\ref{alpham}) holds only for $m=1$ and ${\rm
  Re}(\beta_\l)=-\frac{1}{2}$, which corresponds to the case
$-1<\lambda<1$.  
\end{enumerate}
We note that point (iii) is crucial for Eqn (\ref{Djk}) to give the
correct asymptotic behaviour of $S_n(\ell)$: along the integration contour in
(\ref{Snint}) we always have ${\rm Im}(\lambda)\neq 0$.
Representations with $m\neq 0$ therefore give subleading corrections,
which we are going to analyze in the following section. 

The full result of the generalized Fisher-Hartwig conjecture for the
Toeplitz determinant takes the form
\bea\fl
D_\ell&\sim&(\l+1)^\ell\left(\frac{\l+1}{\l-1}\right)^{-\frac{k_F\ell}\pi}
\sum_{m\in\mathbb{Z}}
L_k^{-2(m+\beta_\l)^2} e^{-2ik_Fm\ell}\nn
\fl&&\hskip 4cm\times \left[G(m+1+\beta_\l)G(1-m-\beta_\l)\right]^2.
\label{Drep2}
\eea
\section{Corrections to the scaling for entanglement}
\label{fhsec}
The leading corrections to scaling for the R\'enyi entropies are obtained
from the ``harmonic'' terms given by the generalized Fisher-Hartwig
conjecture. It follows from (\ref{realpha}) that the most important
corrections arise from the first two contributions with
$m=\pm 1$. Keeping only the three terms corresponding to
$m=-1,0,1$ in (\ref{Drep2}) we obtain the following expression for the
asymptotics of the determinant $D_\ell(\l)$
\bea\
D_\ell
&\sim&
D_\ell^{JK}
\left[1+ e^{-2i k_F \ell} L_k^{-2-4\beta_\l} 
\frac{G^2(2+\beta_\l)G^2(-\beta_\l)}{G^2(1+\beta_\l)G^2(1-\beta_\l)}
\right.\nn\fl && \quad\qquad\left.
+ e^{2i k_F \ell} L_k^{-2+4\beta_\l} 
\frac{G^2(2-\beta_\l)G^2(\beta_\l)}{G^2(1+\beta_\l)G^2(1-\beta_\l)}
\right].
\eea
Here  $D_N^{JK}(\l)$ is given in Eq. (\ref{Djk}). 
Using $G(1+x)/G(x)=\Gamma(x)$ we can rewrite the last formula as
\bea\fl
D_\ell(\l)&\sim&
D_\ell^{JK}\left(1+ \Psi_\ell(\lambda)\right)\ ,\nn\fl
\Psi_\ell(\lambda)&=&
e^{-2i k_F \ell} L_k^{-2(1+2\beta_\l)} 
\frac{\Gamma^2(1+\beta_\l)}{\Gamma^2(-\beta_\l)}+
e^{2i k_F \ell} L_k^{-2(1-2\beta_\l)} 
\frac{\Gamma^2(1-\beta_\l)}{\Gamma^2(\beta_\l)}.
\label{Dgfh}
\eea
It follows from the factorized form of (\ref{Dgfh}) that the
contributions of the correction terms to the entropies are easier
to calculate than the contribution of the leading term $D_\ell^{JK}$
itself. This will enable us to obtain a full analytic answer.  
For large $L_k$ we have (we recall that $d_n(\ell)=S_n(\ell)-S_n^{JK}(\ell)$)
\bea
\fl d_n(\ell)&\sim& \frac1{2\pi i}\oint d\lambda\ e_n(\l) \frac{d\ln 
\left[1+ \Psi_\ell(\lambda)\right]}{d\l} 
=\frac1{2\pi i}\oint d\lambda\ e_n(\l) \frac{d
\Psi_\ell(\lambda)}{d\l}+\ldots.
\label{approx}
\eea
The contour integral can be written as the sum of two contributions
infinitesimally above and below the interval $[-1,1]$ respectively, i.e. 
\bea
d_n(\ell)&\sim&
\frac1{2\pi i}\left[\int_{-1+i\epsilon}^{1+i\epsilon}-
  \int_{-1-i\epsilon}^{1-i\epsilon}\right]d\lambda\
 e_n(\l) 
 \frac{d\Psi_\ell(\lambda)}{d\l} .
\label{int2}
\eea
This shows that we only require the discontinuity across the branch
cut. The only discontinuous function is $\beta_\l$, which for 
$-1<x<1$ behaves as
\be
\beta_{x\pm i\epsilon}= -i w(x) \mp \frac12\,, \qquad {\rm with}\qquad 
w(x)=\frac1{2\pi} \ln \frac{1+x}{1-x}\,.
\ee
We now change variables from $\lambda$ to $w$
\be
\l= \tanh(\pi w)\ ,\quad -\infty<w<\infty.
\ee
We have
\bea\fl
\left[L_k^{-2-4\beta} \frac{\Gamma^2(1+\beta)}{\Gamma^2(-\beta)}\right]_{\beta=-i w-\frac12}-
\left[L_k^{-2-4\beta} \frac{\Gamma^2(1+\beta)}{\Gamma^2(-\beta)}\right]_{\beta=-i w+\frac12}&\simeq&
L_k^{4 i w} \gamma^2(w),
\nn \fl
\left[L_k^{-2+4\beta} \frac{\Gamma^2(1-\beta)}{\Gamma^2(\beta)}\right]_{\beta=-iw-\frac12}-
\left[L_k^{-2+4\beta} \frac{\Gamma^2(1-\beta)}{\Gamma^2(\beta)}\right]_{\beta=-iw +\frac12}&\simeq&
-L_k^{-4 i w} \gamma^2(-w), 
\nonumber
\eea
where we have dropped terms of order $O(L_k^{-4})$ compared to the
leading ones and we have defined
\be
\gamma(w)=\frac{\Gamma(\frac12-i w)}{\Gamma(\frac12+i w)}.
\ee
Integrating by parts and using
\be
\frac{d}{dw}e_n(\tanh(\pi w))=
\frac{\pi n}{1-n}(\tanh(n\pi w)-\tanh(\pi w))\,,
\label{intfin}
\ee
we arrive at
\bea\fl
d_n(\ell)&\sim&\frac{i n}{2(1-n)}\int_{-\infty}^\infty dw 
(\tanh(\pi w)-\tanh(n\pi w))
\times\nn\fl &&\quad\quad\quad\times
 \left[e^{-2ik_F \ell} 
L_k^{4 i w} \gamma^2(w)
-e^{2ik_F \ell} 
L_k^{-4 i w} \gamma^2(-w)
\right]+\ldots
\label{finint}
\eea
For large $\ell$ the leading contribution to the integral arises from
the poles closest to the real axis. These are located at
$w_0=i/2n$ ($w_0=-i/2n$) for the first (second) term in (\ref{finint}).
Evaluating their contributions to the integral gives
\be
\fl d_n(\ell)=\frac{ 2 \cos(2 k_F \ell) }{1-n} (2\ell
|\sin k_F |)^{-2/n} 
\left[\frac{\Gamma(\frac{1}{2}+\frac{1}{2n})}
{\Gamma(\frac{1}{2}-\frac{1}{2n})}\right]^2
+o\big(\ell^{-2/n}\big).
\label{Snfinal}
\ee
This result implies that at half-filling ($k_F=\frac{\pi}{2}$) and
$n>1$ the corrections are positive (negative) for odd (even) $\ell$.

\subsection{Subleading Corrections}

Eqn (\ref{Snfinal}) describes the asymptotic behaviour in the limit
$L_k\to\infty$, $n$ fixed. It provides a good approximation for
large, finite $\ell$ as long as $\ln( L_k)\gg n$. This is a strong restriction
already for moderate values of $n$. For example, $L_k$ is required to be
larger than $10^4$ for $n=10$. For practical purposes it is useful to
know the corrections to $S_n(\ell)$ for large $\ell$ but $\ln( L_k)$
not necessarily much larger than $n$. 
In this regime there are two main sources of corrections to
(\ref{Snfinal}). 
\begin{enumerate}
\item{}
The integral (\ref{intfin}) is no longer dominated by the poles
closest to the real axis and contributions from further poles need to
be included. These give rise to corrections proportional to $L_k^{-2 q
  /n}$, with $q$ integer. 
\item{} Further terms in the expansion of the logarithm in
Eqn (\ref{approx}) need to be taken into account. The corresponding
contributions are proportional to $e^{\pm i 2p k_F \ell}$ with
$p=2,3,\dots$.

At half-filling (zero magnetic field) the situation is different in
that terms with odd $p$ all give rise to an overall factor $(-1)^\ell$
and hence modify the staggered contribution to $S_n(\ell)$,
while terms with even $p$ add to the smooth (non-oscillatory) part
already present in $S_n^{JK}(\ell)$.
\end{enumerate}
We now take both types of corrections into account. We first consider
the series expansion of the logarithm in Eq. (\ref{approx})  
\be
\ln\big[1+\Psi_\ell(\lambda)\big]=\sum_{p=1}^\infty 
\frac{(-1)^{p+1} \big(\Psi_\ell(\lambda)\big)^p}{p}\,.
\ee
Recalling the explicit expression (\ref{Dgfh}) for
$\Psi_\ell(\lambda)$ leads to a binomial sum
\bea
\big(\Psi_\ell(\lambda)\big)^p
&=&\left(e^{-2i k_F\ell}L_k^{-2(1+2\beta_\l)}c_{\beta_\l} + 
e^{2i k_F\ell}L_k^{-2(1-2\beta_\l)}c_{-\beta_\l}\right)^p\nn &=&
\sum_{q=0}^p {p \choose q} e^{2ik_F \ell (2q-p)} L_k^{-2p} L_k^{-4(p-2q)\beta_\l}c_{\beta_\l}^{p-q}
c_{-\beta_\l}^{q}\,,
\eea
where we have introduced the shorthand notation
$c_{\beta}=(\Gamma(1+\beta)/\Gamma(-\beta))^2$. When calculating the
discontinuity across the branch cut running from $\lambda=-1$ to
$\lambda=1$ all terms other than $q=0$ and $q=p$ give rise to terms
that are subleading in $L_k$. Hence we may approximate 
\bea
\fl
\big(\Psi_\ell(\tanh(\pi w)+i\epsilon)\big)^p-
\big(\Psi_\ell(\tanh(\pi w)-i\epsilon)\big)^p&\approx&
e^{-2ik_F \ell p}L_k^{4iwp}c_{-iw-1/2}^p\nn
&&
+e^{2ik_F \ell p}L_k^{-4iwp}c_{-iw+1/2}^p\,.
\eea
The analog of (\ref{finint}) then reads
\bea\fl
d_n(\ell)&\sim&\sum_{p=1}^\infty\frac{(-1)^{p+1}}{p}
\frac{i n}{2(1-n)}\int_{-\infty}^\infty dw 
(\tanh(\pi w)-\tanh(n\pi w))\nn\fl  &\times&
\!\left[e^{-2ipk_F \ell}  L_k^{4 i w p} \gamma^{2p}(w)
-e^{2ipk_F \ell}  L_k^{-4 i wp} \gamma^{2p}(-w)
\right].
\label{finint2}
\eea
The integral is carried out by contour integration, taking the two
terms in square brackets into account separately. The first (second)
contribution has simple poles in the upper (lower) half plane at
$w_q= i\frac{2q-1}{2n}$ ($w_q= -i\frac{2q-1}{2n}$), where $q$ is a
positive integer such that $2q-1\neq n,3n,5n,\ldots$.
Contour integration then gives
\bea
\fl
d_n(\ell)&=&\frac{2}{1-n}{\sum_{p,q=1}^\infty}
\frac{(-1)^{p+1}}{p}
\cos(2 k_F \ell p)
L_k^{-\frac{2p(2q-1)}{n}}
\big(Q_{n,q}\big)^{p}
+{\cal O}(L_k^{-1-2/n}),
\label{main}
\eea
where the constans $Q_{n,q}$ have been defined in (\ref{Qn}).
In the sum over $q$ the special values $2q-1\neq n,3n,5n,\ldots$ are
to be omitted. In particular, this means that for $n=1$  all these
corrections are absent. 
Eqn (\ref{main}) is one of the main results of our work. It shows that
there are contributions to the R\'enyi entropies with oscillation
frequencies that are arbitrary multiples of $2k_F$.

At half-filling ($k_F=\pi/2$) certain simplifications occur. 
For even $\ell$ we find
\be\fl
d_n(\ell)\sim
\frac{2}{1-n}\left[
(2\ell)^{-\frac{2}{n}}Q_{n,1} -(2\ell)^{-\frac{4}{n}}
\frac{Q_{n,1}^{2}}{2}
 +(2\ell)^{-\frac{6}{n}}\left(\frac{Q_{n,1}^3}{3} +Q_{n,3}\right)
\right]+\dots\,,
\ee
while for odd $\ell$ we obtain
\be\fl
d_n(\ell)\sim
\frac{-2}{1-n}\left[ (2\ell)^{-\frac{2}{n}} Q_{n,1}
+(2\ell)^{-\frac{4}{n}} \frac{Q_{n,1}^{2}}{2}
+(2\ell)^{-\frac{6}{n}}\left(\frac{Q_{n,1}^{3}}{3} +Q_{n,3} \right)
\right]+\dots\,.
\ee
In all the above analysis we have ignored contributions to the
generalized Fisher-Hartwig conjecture with $|m|>1$. While these lead
to oscillatory contributions with frequencies that are integer
multiples of $2k_F$ they are suppressed by additional powers of
$\ell^{-1}$ and hence are subeading, even in the case where $n$ is not
small. 

It is apparent from (\ref{main}) that the limit $n\to\infty$ deserves
special attention. $S_\infty(\ell)$ is known in the literature as {\it
  single copy entanglement} \cite{sce}.   
Here it is neccessary to sum up an infinite number
of contributions in order to extract the large-$\ell$ asymptotics.
We note that the large-$n$ limit is not only of academic interest, but
will provide information on the behaviour of $S_n(\ell)$ in the regime
$n\gg \ln L_k$, $L_k\gg 1$. 
\subsection{Large $n$ limit of $S_n(\ell)$}
In order to investigate the limit $n\to\infty$ we consider
eqn (\ref{finint2}), but now first take the parameter $n$ to infinity
and then carry out the resulting integrals. This gives
\bea\fl
d_\infty(\ell)&\sim&
\frac{i}2\sum_{p=1}^\infty \frac{(-1)^p}{p}
\int_{-\infty}^\infty dw (\sgn(w)-\tanh(\pi w))\nn 
\fl&&\times
\left[e^{-2ik_F \ell p} 
L_k^{4 i wp} \left[\gamma(w)\right]^{2p}
e^{2ik_F \ell p} 
L_k^{-4 i wp}\left[
\gamma(-w)\right]^{2p}
\right]\nn\fl
&=&-\sum_{p=1}^\infty \frac{(-1)^p}{p}\left[
e^{-2ik_F \ell p}  {\rm Im} \int_{0}^\infty dw 
[1-\tanh(\pi w)]
L_k^{4 i wp} \left[\gamma(w)\right]^{2p}
\right. \nn\fl  
&&\left. \qquad\qquad\quad-
e^{2ik_F \ell p}  {\rm Im} \int_{0}^\infty dw [1-\tanh(\pi w)]
L_k^{-4 i wp} \left[
\gamma(-w)\right]^{2p}
\right].
\eea
Using that the first singularity in the upper (lower) half plane
occurs at $w=i/2$ ($w=-i/2$) we deform the contours to run parallel to
the real axis with imaginary parts $i/4$ and $-i/4$ respectively,
i.e. for the first term we use
$$
\int_0^\infty dw\ f(w)= \int_0^{i/4} dw\ f(w) +\int_{i/4}^{\infty+i/4}
dw\ f(w).
$$
It is straightforward to show that the second integral contributes
only to order $O(1/L_k)$ and does not give rise to logarthimic corrections.
Hence the leading contribution is of the form
\bea\fl
d_\infty(\ell)&\sim&\sum_{p=1}^\infty \frac{(-1)^p}{p}
\left[e^{-2ik_F \ell p}  {\rm Re} \int_{0}^{1/4} dz
(1-i\tan(\pi z))
L_k^{-4 z p} \left(\gamma(iz)\right)^{2p}
\right.\nn\fl &&\qquad \left. +
e^{2ik_F \ell p} {\rm Re} \int_{0}^{1/4} dz(1+i\tan(\pi z))
L_k^{-4 z p} \left(\gamma(iz)\right)^{2p}
\right]\nn\fl &=&
2\sum_{p=1}^\infty \frac{(-1)^p}{p} \cos(2k_F \ell p)
\int_{0}^{1/4} dz e^{-4 z p \ln L_k} 
\left(\gamma(iz)\right)^{2p}\ .
\eea
For large $L_k$ the dominant contribution to integral is obtained by
expanding $\left(\gamma(iz)\right)^{2p}$ in a power series around $z=0$
\bea\fl
d_\infty(\ell)&\sim&
2\sum_{p=1}^\infty \frac{(-1)^p}{p} \cos(2k_F \ell p)
\int_{0}^{1/4} dz e^{-4 z p \ln L_k} (1+ 4 p z \Psi(1/2)+\dots)\nn\fl
&=&
2\sum_{p=1}^\infty \frac{(-1)^p}{p^2} \cos(2k_F \ell p)
\left[\frac1{4\ln L_k}+ \frac{ \Psi(1/2)}{4\ln^2 L_k}+\dots\right]
+O(L_k^{-1}),
\label{expansion}
\eea
where $\Psi(x)$ is the digamma function. The leading contribution can
be expressed in terms of the dilogarithm function ${\rm Li}_2(x)$ using
\be
2\sum_{p=1}^\infty \frac{(-1)^p}{p^2} \cos(2k_F \ell p)=
{\rm Li}_2(-e^{i2k_F \ell})+{\rm Li}_2(-e^{-i2k_F \ell}).
\ee
In the half-filled case ($k_F=\pi/2$) our result takes
a particularly simple form
\be
d_\infty(\ell)\sim
\frac1{2\ln L_k}\sum_{p=1}^\infty \frac{(-1)^{p(\ell+1)}}{p^2}=
\cases{\displaystyle
\frac1{2\ln L_k} \frac{\pi^2}6 & $\ell$ {\rm odd} \,,\\ \displaystyle
-\frac1{2\ln L_k} \frac{\pi^2}{12} & $\ell$ {\rm even}\,.
}
\label{logcorr}
\ee
Summing some of the subleading terms in (\ref{expansion}) to all
orders in $(\ln(L_k))^{-1}$ leads to an expression of the form
\be
d_\infty(\ell)\sim
\frac{\pi^2}{24\ln(b L_k)}
\cases{\displaystyle
2 & $\ell$ {\rm odd} \,,\\ \displaystyle
-1 & $\ell$ {\rm even}\,,
}
\label{Sinfty}
\ee
where $b=\exp(-\Psi(1/2))\approx 7.12429$. This is found to be in good
agreement with numerical computations.

\section{Connection with random matrix theory}
\label{rmsec}
Keating and Mezzadri \cite{km-05} have shown that the Toeplitz
determinant $D_\ell(\l)$ is related to an important quantity in random
matrix theory, namely the gap probability for the circular unitary
ensemble (CUE). The generating function 
$E^{\rm CUE}_\ell [(0,\phi);\xi]$ (in the following we will drop all
arguments to ease notations) for the probability of finding 
exactly $k$ eigenvalues $e^{i\theta} $ within the segment
$ \theta \in (\pi-\phi,\pi] $ of the unit circle is given by
\cite{fw-05}
$$
 E^{\rm CUE}_\ell \equiv{1 \over  (2\pi)^\ell \ell! }
 \left( \int^{\pi}_{-\pi} - \xi\int^{\pi}_{\pi-\phi} \right) d\theta_1 \ldots
 \left( \int^{\pi}_{-\pi} - \xi\int^{\pi}_{\pi-\phi} \right) d\theta_\ell
 \prod_{1 \leq j < k \leq \ell} |e^{i\theta_j}-e^{i\theta_k}|^2 .
$$
This is equal to the determinant of the Toeplitz matrix \cite{fw-05}
\begin{equation}
W_{ij}=w_{i-j},\qquad {\rm with}\;\;   w_n = \delta_{n,0} + {\xi \over 2\pi i}(-1)^{n+1}{e^{in\phi}-1 \over n} .
\end{equation}
It then follows that for $\xi=2/(\l+1)$ and $\phi=2 k_F$ we have
\cite{km-05}
\be
D_\ell(\l)=(\l+1)^\ell E^{\rm CUE}_\ell\,.
\ee
For any value of $\ell$ the generating function $E^{\rm CUE}_\ell$ 
can be determined from a recurrence relations connected to the
Painlev\'e VI transcendent \cite{fw-05}. The recurrence relation reads 
\be\fl
   x_{\ell}x_{\ell-1} - c
 = {1-x^2_\ell \over 2 x_\ell}\left[ (\ell+1)x_{\ell+1}+(\ell-1)x_{\ell-1} \right] 
    - {1-x^2_{\ell-1} \over 2 x_{\ell-1}}\left[\ell x_{\ell}+(\ell-2)x_{\ell-2} \right] ,
    \label{recx}
\ee
where $c=\cos k_F$ and the initial values are
\begin{equation}
   x_{-1} = 0, \quad x_{0} = 1, \quad
   x_{1} = -{\xi \over \pi}{\sin k_F \over 
                            1 - \frac{\xi}{\pi}k_F } .
\end{equation} 
The generating function is related to $x_\ell$ by
\begin{equation}
 {E^{\rm CUE}_{\ell+1}E^{\rm CUE}_{\ell-1} \over (E^{\rm CUE}_{\ell})^2} = 1 - x^2_\ell =
 \frac{D_{\ell+1} D_{\ell-1}}{D_\ell^2}.
 \label{recr}
\end{equation}
For the sake of completeness we quote the values of the generating
function for $\ell=0$ and $\ell=1$
\begin{equation}
  E^{\rm CUE}_{0} = 1, \quad E^{\rm CUE}_{1} = 1 - {\xi \over 2\pi}\phi .
\end{equation}
\subsection{Leading large-$\ell$ asymptotics of $x_\ell$}
In Ref. \cite{km-05} it was suggested to combine the asymptotic results
(\ref{Djk}) following from the Fisher-Hartwig conjecture with the
recurrence relation (\ref{recx}) in order to obtain further corrections to the
large-$\ell$ behaviour of the R\'enyi entropies. Inserting (\ref{Djk})
into (\ref{recr}) suggests that \cite{km-05}
\be
x_\ell= \frac{\sqrt2 |\beta_\l|}{\ell}+O(\ell^{-2})\,. 
\label{wrongkm}
\ee
However, a numerical solution of the recurrence relation shows that
(\ref{wrongkm}) does not generally provide the correct large-$\ell$
asymptotics of $x_\ell$. The reason for this is as follows. When we
substitute the ``full'' result (\ref{Dgfh}) of the \emph{generalized}
Fisher-Hartwig conjecture into (\ref{recx}) we find that the
contributions due to the representations with $m=\pm1$ behave as
$\ell^{-1\pm   2\beta_\l}$ for large $\ell$. For any ${\rm
  Re}(\beta_\l)\neq 0$ one of these will dominate over the
contribution arising from the $m=0$ term that gives rise to
Eq. (\ref{wrongkm}). In other words \emph{subleading} contributions to
$D_\ell(\lambda)$ give rise to the \emph{leading} large-$\ell$
behaviour of $x_\ell$!  

\begin{figure}[t]
\begin{center}
\includegraphics[width=\textwidth]{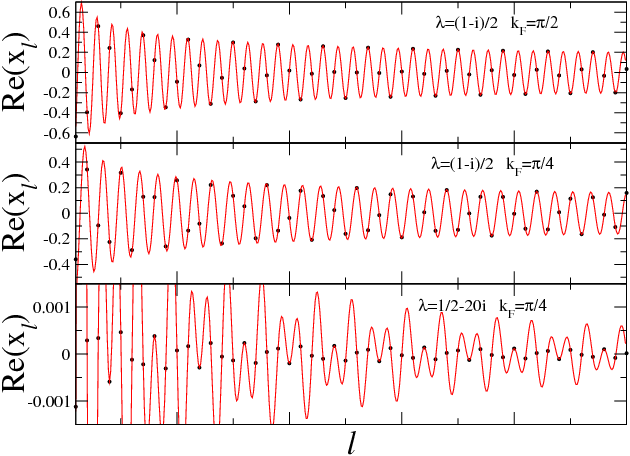}
\end{center}\caption{Real part of $x_\ell$ as a function of $\ell$ for
several values of $\lambda$ and $k_F$. The points are obtained from a
numerical solution of the recurrence relation (\ref{recx}). The
continuous lines are obtained from the asymptotic prediction
(\ref{xasy2}) by the replacement $(-1)^\ell\to e^{i\pi \ell}$. The
first two panels correspond to the same value $\beta_\l\simeq 0.323792 -
0.128075 i$ but two different values of $k_F$. The last panel corresponds
to $\beta_\l\sim 0.0158924 - 0.0003966 i$ and hence the contributions
of both terms in Eq. (\ref{xasy2}) are important. In all cases
we observe good agreement of the theoretical prediction (\ref{xasy2})
with the numerical data.} 
\label{fig:xN}
\end{figure}

We now show in more detail how to extract the large-$\ell$
behaviour of $x_\ell$ from that of $D_\ell(\lambda)$. In order to keep
things simple, we focus on the case ${\rm Re}(\beta_\l)>0$. Here we may
neglect the terms with $|m|>1$ and $m=-1$ in (\ref{Drep2}), which
leads to
\bea
\frac{D_{\ell+1}D_{\ell-1}}{D_\ell^2}\sim
\left[1+\frac{2\beta_\l^2}{\ell^2}\right]
\left(1-
4a_0^2 e^{2i k_F \ell}\ell^{-2+4\beta_\l}\sin^2 k_F\right)+\dots,
\label{Dratio}
\eea
where we have introduced 
\be
a_0=(2\sin k_F)^{-1+2\beta_\l}\frac{\Gamma(1-\beta_\l)}{\Gamma(\beta_\l)}.
\label{a0}
\ee
The contribution in square brackets arises from the $m=0$
Fisher-Hartwig term and is the result quoted in Ref. \cite{km-05}. 
For ${\rm Re}(\beta_\l)>0$ this term is subleading and we obtain instead
\be
x_\ell\sim
(-1)^\ell e^{i k_F \ell} \ell^{-1+2\beta_\l}(2\sin k_F)^{2\beta}
\frac{\Gamma(1-\beta_\l)}{\Gamma(\beta_\l)}\ ,\
{\rm Re}(\beta_\l)>0.
\label{xasy}
\ee
Here we have fixed the sign of $x_\ell$ by requiring that the
expression (\ref{xasy}) asymptotically satisfies the recurrence
relation (\ref{recx}). The analogous analysis in the case
${\rm Re}(\beta_\l)<0$ gives
\be
x_\ell\sim
(-1)^\ell e^{-i k_F \ell} \ell^{-1-2\beta_\l}
(2\sin k_F)^{-2\beta_\l} \frac{\Gamma(1+\beta_\l)}
{\Gamma(-\beta_\l)}\ ,\
{\rm Re}(\beta_\l)<0.
\label{xasy1}
\ee
We may combine Eqns (\ref{xasy}) and (\ref{xasy1}) into a single
equation 
\be\fl
x_\ell\sim\frac{(-1)^\ell}{\ell} \left[e^{i k_F \ell} (2\ell\sin
  k_F)^{2\beta_\l} \frac{\Gamma(1-\beta_\l)}{\Gamma(\beta_\l)}+
e^{-i k_F \ell} (2\ell\sin k_F)^{-2\beta_\l} \frac{\Gamma(1+\beta_\l)}{\Gamma(-\beta_\l)}\right]+\dots
\label{xasy2}
\ee
We emphasize that (\ref{xasy2}) must not be understood as giving the
two leading terms in the large-$\ell$ asymptotic expansion of $x_\ell$
because e.g. for ${\rm Re}(\beta_\l)>\frac{1}{6}$
there are other contributions to $x_\ell$ that decay more slowly than
$\ell^{-1-2\beta_\l}$. 
In order to check the result (\ref{xasy2}) we have
solved the recurrence relation (\ref{recx}) numerically for a number
of different values of $\lambda$ and $k_F$. In Fig. \ref{fig:xN} we
compare the asymptotic expression (\ref{xasy2}) against
the numerically computed values for $x_\ell$. The agreement is seen to
be excellent in all cases.

\subsection{Asymptotic expansion for $x_\ell$ and analytic corrections
to the gFHC expression for $D_\ell(\lambda)$} 
We now turn to the derivation of contributions to the large-$\ell$
asymptotic expansion for $D_\ell(\lambda)$ that are not contained in the
gFHC. This will be achieved by utlizing the recurrence relation
(\ref{recx}). 

We expect the asymptotics of $D_\ell(\lambda)$ to be such that
each harmonic term predicted by gFHC is multiplied by a function analytic
in $1/\ell$. Restricting out attention to the first three
harmonic terms (i.e. $m=-1,0,1$), this leads to an expansion of the form
\bea\fl
\frac{D_\ell}{D_\ell^{JK}}&\sim&
\left[1+ \frac{c_1}{\ell}+ \frac{c_2}{\ell^2}+\ldots\right]
+ a_0^2 e^{2i k_F \ell} \ell^{-2(1-2\beta_\l)} \left[1+ \frac{a_1}{\ell}
  +\frac{a_2}{\ell^2}+\dots \right]\nn\fl&&\quad 
+ b_0^2 e^{-2i k_F \ell} \ell^{-2(1+2\beta_\l)}  
\left[1+ \frac{b_1}{\ell}+ \frac{b_2}{\ell^2}+\dots \right] ,
\label{DexpT}
\eea
where $a_0$ is given by (\ref{a0}) and
\be
b_0= (2\sin k_F)^{-1-2\beta_\l} \frac{\Gamma(1+\beta_\l)}{\Gamma(-\beta_\l)}\,.
\ee
We note that by definition we have
\be
(2\sin k_F)^2 a_0 b_0=-\beta_\l^2\,.
\label{a0b0}
\ee
We now proceed in a straightforward albeit extremely tedious way:
\begin{enumerate}
\item{}
We first insert (\ref{DexpT}) into (\ref{recr}) in order to obtain an
expression for the asymptotic expansion for $x_\ell$.
\item{}
We input the resulting expression into the recurrence relation
(\ref{recx}) for $x_\ell$ and determine the parameters characterizing
the asymptotic expansion of $x_\ell$ order by order in $\ell^{-1}$.
\end{enumerate}

The result of step (i) for ${\rm Re}(\beta_\l)>0$ is
\bea
\frac{x_\ell}{x_\ell^{\rm asy}}&=&
\sum_{j=0}^3(-1)^j
\left[
1+\frac{o_{j1}}\ell +\frac{o_{j2}}{\ell^{2}}+\frac{o_{j3}}{\ell^{3}}+
\ldots\right]a_0^{2j}\ e^{2ik_Fj\ell}\ \ell^{j(-2+4\beta_\l)}
\nn
&+&\sum_{j=1}^2
\frac{\ell^{-4j\beta_\l} e^{-2ik_F j\ell}}{(2 a_0 \sin k_F)^{2j}} 
\left[q_{j0}+\frac{q_{j1}}\ell+\frac{q_{j2}}{\ell^2} +\cdots\right]
+\ldots,
\label{Dexp}
\eea
where we have written only the terms required for our purposes and where
have introduced the quantity 
\be
x_\ell^{\rm asy}=(-1)^\ell e^{i k_F \ell} \ell^{-1+2\beta_\l} 2 a_0
\sin k_F .
\ee
The explicit expressions for the coefficients $o_{jl}$ and
$q_{jl}$ in terms of the expansion coefficients $a_j,b_j$, and $c_j$ 
characterizing the large-$\ell$ asymptotics of $D_\ell$ are reported
in \ref{appA}. 

Step (ii) consists of substituting (\ref{Dexp}) in the recurrence
relation (\ref{recx}) and determining the coefficients $o_{jl}$ and
$q_{jl}$. The non-linearity of (\ref{recx}) renders this a
very difficult task, because terms at different orders in $\ell^{-1}$
in $x_\ell$ contribute to the same order in the recurrence relation. For
this reason it is crucial to retain sufficiently many terms in
(\ref{Dexp}). The results of this procedure are reported
in \ref{appB}.

Combining the results reported in \ref{appA} and \ref{appB} then
yields the desired expressions for the expansion coefficients
$a_j,b_j$ and $c_j$
\bea
\label{finalcoef}
c_1(\beta_\l)&=&2  \beta_\l^3i \cot k_F\,,\nn
c_2(\beta_\l)&=& \frac{\beta_\l^2}6 (-1 + 7 \beta_\l^2 + 12 \beta_\l^4
- 3 \beta_\l^2 (5 + 4 \beta_\l^2) \csc^2 k_F) \,,\nn
b_j(\beta_\l)&=&c_j(\beta_\l+1)\ ,\quad j=1,2,\nn
a_j(\beta_\l)&=&c_j(\beta_\l-1)\ ,\quad j=1,2.
\eea

\section{Corrections to the von Neumann entropy}
\label{vnsec}
Having determined the asymptotic expansion for $D_{\ell}(\lambda)$ we
may now use (\ref{Snint}) to calculate additional subleading
contributions to the R\'enyi entropies.
We first consider the von Neumann entropy (the case $n=1$), in which
as we have seen above all harmonic contributions vanish.
Taking the limit $n\to 1$ in (\ref{Snint}) and following through the
same steps as in section \ref{fhsec} we find
\bea\fl
d_1(\ell)&\sim&\frac{i }{2}\int_{-\infty}^\infty dw \frac{\pi w}{\cosh^2\pi w} 
\left[\frac{c_1^+ - c_1^-}{\ell}
+\frac{2c_2^+ - (c_1^+)^2-2c_2^- +(c_1^-)^2}{2\ell^2} \right]\,,
\label{vNE}
\eea
where we have defined 
\be
c_j^{\pm}=c_j(-i w\mp \frac12)\ ,\quad j=1,2.
\label{cjpm}
\ee
Here $c_{1,2}$ are given by (\ref{finalcoef}) and we have used 
\be
\lim_{n\to1}\frac{\tanh(\pi w)-\tanh(n\pi w)}{1-n}= \frac{\pi
  w}{\cosh^2\pi w} \,. 
\ee
As $c_1^+-c_1^-$ is an even function of $w$ the ${\cal O}(\ell^{-1})$
contribution in (\ref{vNE}) vanishes. The ${\cal O}(\ell^{-2})$
contribution can be calculate analytically using the integrals
\be
\int_{-\infty}^\infty dw \frac{\pi w^2}{\cosh^2\pi w}=\frac16\,,  \qquad
\int_{-\infty}^\infty dw \frac{\pi w^4}{\cosh^2\pi w}=\frac{7}{120}\,,
\ee
which gives the final result
\be
d_1(\ell)\sim-\frac1{12\ell^2}\left[\frac15+\cot^2 k_F\right]\,.
\ee
The simplicity of this answer suggests the existence of a much more
straightforward derivation than ours.

\section{Corrections to the R\'enyi entropies}
\label{Rnsec}
The case of the R\'enyi entropies is more complicated because the
contribution of the harmonic terms does not vanish. Our starting point
is the expansion (\ref{DexpT}) for the Toeplitz determinant, which we
express in the form
\bea
\frac{D_\ell}{D_\ell^{JK}}&\sim&
1+\Psi_\ell(\beta_\l)+\frac{\delta\Psi^{(1)}_\ell(\beta_\l)}{\ell}
+\frac{\delta\Psi^{(2)}_\ell(\beta_\l)}{\ell^2}+\ldots
\label{DexpT2}
\eea
Here $\Psi_\ell(\beta_\l)$ are the contributions we have taken into
account previously in section \ref{fhsec}. The logarithm of the
Toeplitz determinant is expanded as
\bea\fl
\ln\left[ \frac{D_\ell}{D_\ell^{JK}}\right]&\sim&
\ln\left[1+\Psi_\ell(\beta_\l)
+\frac{\delta\Psi^{(1)}_\ell(\beta_\l)}{\ell} 
+\frac{\delta\Psi^{(2)}_\ell(\beta_\l)}{\ell^2}\right]\nn\fl
&\approx&\sum_{p=1}^\infty\frac{(-1)^{p+1}}{p}
\Bigg\{\left[\Psi_\ell(\beta_\l)\right]^p+p\frac{\delta\Psi_\ell^{(1)}(\beta_\l)
\left[\Psi_\ell(\beta_\l)\right]^{p-1}}{\ell}\nn
\fl&&\qquad\qquad
+p\frac{[\Psi_\ell(\beta_\l)]^{p-1}\delta\Psi_\ell^{(2)}(\beta_\l)+\frac{p-1}{2}
\left[\delta\Psi_\ell^{(1)}(\beta_\l)\right]^2
\left[\Psi_\ell(\beta_\l)\right]^{p-2} }{\ell^2}\Bigg\}\nn
\fl&\equiv&
\chi^{(0)}(\beta_\l)+\frac{\chi^{(1)}(\beta_\l)}{\ell}
+\frac{\chi^{(2)}(\beta_\l)}{\ell^2}.
\eea
Following through the same steps as in section \ref{fhsec} we then
arrive at the following expansion for the R\'enyi entropies
\bea\fl
d_n(\ell)&\sim&\frac{i n}{2(1-n)}\int_{-\infty}^\infty dw 
(\tanh(\pi w)-\tanh(n\pi w))
\nn\fl &&\qquad\qquad\times\left[
\chi^{(0)}(\beta_\l)+\frac{\chi^{(1)}(\beta_\l)}{\ell}
+\frac{\chi^{(2)}(\beta_\l)}{\ell^2}\right]_{\beta_\l=-iw+\frac{1}{2}}^{\beta_\l=-iw-\frac{1}{2}}\nn\fl
&=&d_n^{(0)}(\ell)+d_n^{(1)}(\ell)+d_n^{(2)}(\ell).
\eea
The contribution $d_n^{(0)}(\ell)$ has been determined in
section \ref{fhsec}.
The other two contributions are calculated by the same method as in
section \ref{fhsec} and we find
\bea\fl
d_n^{(1)}(\ell)&\sim&
\frac{ 2 \cos(k_F ) }{1-n}
\sum_{p,q=1}^\infty(-1)^{p+1}\sin(2k_Fp\ell)\
L_k^{-1-\frac{2p(2q-1)}{n}}\nn\fl
&&\qquad\quad\times\left[1+3\left(\frac{2q-1}{n}\right)^2\right]
\left[\frac{\Gamma\Big(\frac{1}{2}+\frac{2q-1}{2n}\Big)}
{\Gamma\Big(\frac{1}{2}-\frac{2q-1}{2n}\Big)}\right]^{2p}\ ,
\label{dS1}\\
\fl
d_n^{(2)}(\ell)&\sim&
\frac{2}{n-1}
\sum_{p,q=1}^\infty(-1)^{p}
L_k^{-2-\frac{2p(2q-1)}{n}}
\left[\frac{\Gamma\Big(\frac{1}{2}+\frac{2q-1}{2n}\Big)}
{\Gamma\Big(\frac{1}{2}-\frac{2q-1}{2n}\Big)}\right]^{2p}
\Big[
B^{(n)}_{p,q}e^{2ipk_F\ell}+{\rm h.c.}\Big]\nn
\fl&&
+\frac{1}{\ell^2}\frac{n+1}{1440 n^3}\left(49-n^2+\frac{15(3n^2-7)}
{\sin^2k_F}\right).
\label{dS2}
\eea
Explicit expressions for the coefficients $B^{(n)}_{pq}$ are given in
\ref{app:C}.

\section{Numerical results}
\label{numsec}
Given our asymptotic expansion a natural question to ask is how well
it approximates the R\'enyi entropies for large but finite block
lengths $\ell$. In order to address this question we will now present
a number of comparisons between our asymptotic result and numerically
exact expressions for $S_n(\ell)$. The latter are obtained by
determining the eigenvalues of the Toeplitz matrix $C_{nm}$ in
Eq. (\ref{Cnm}) and computing $S_n$ from Eq. (\ref{Sn}). 
\subsection{Leading contributions to $d_n(\ell)$}
\begin{figure}[t]
\begin{center}
\includegraphics[width=0.8\textwidth]{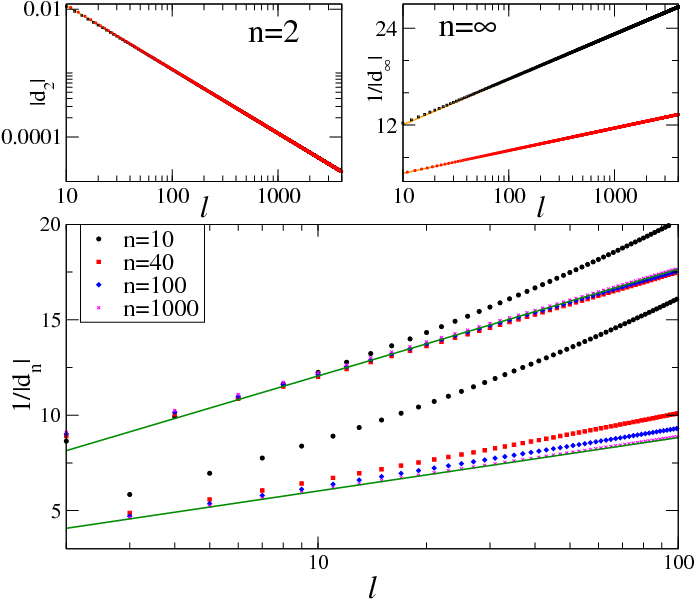}
\end{center}\caption{
Top: $d_n(\ell)=S_n(\ell)-S_n^{JK}(\ell)$ at half-filling
($k_F=\frac{\pi}{2}$) for $n=2$ and $n=\infty$ compared to the the
asymptotic expressions (straight lines for even/odd $\ell$
respectively). The agreement is seen to be 
excellent even for moderate values of $\ell$.
Bottom: $|d_n(\ell)|^{-1}$ as a function of $\ell$ for several values
of $n$ and $k_F=\frac{\pi}{2}$. The straight lines show the asymptotic
results (\ref{Sinfty}) in the limit $n\to\infty$ for even and odd
$\ell$ respectively. We see that for large $n$ the correction
$d_n(\ell)$ exhibits a logarithmic increase up to a block size $\ln
\ell\sim n$, when the asymptotic behaviour starts to be seen (as we
are plotting $|d_n(\ell)|^{-1}$ the asymptotic behaviour corresponds
to a $\ell^{2/n}$ power-law increase with $\ell$). }
\label{fig:Rn}
\end{figure}
In the top two panels of Fig.\ref{fig:Rn} we plot the absolute value
of $d_n(\ell)$
for $n=2$ and $n=\infty$ at $k_F=\pi/2$ and compare it to the
leading asymptotic expressions (\ref{Sncorrintro}) and (\ref{Sinfty})
respectively. We see that the asymptotic expressions give good
agreement with the numerically exact results even for moderate values
of $\ell$. 

The next issue we turn to is the behaviour of $S_n(\ell)$ for large, finite
values of $n$. In this case the asymptotic power-law
$d_n(\ell)\propto\ell^{-2/n}$ only emerges for very large block
lengths $\ln \ell\gg n$. On the other hand, for smaller values of
$\ell$ the numerical data is seen to follow the large $n$ prediction
(\ref{Sinfty}) as is shown in the bottom panel of
Fig. \ref{fig:Rn}. Here we plot $1/|d_n(\ell)|$, which at large values 
of $\ell$ will grow as $\ell^{2/n}$. A logarithmic behaviour for small
small values of $\ell$ is clearly visible, which then crosses over to
the expected $\ell^{2/n}$ regime at approximately $\ln \ell\sim n$.
It appears that the crossover scale is larger for even $\ell$.
We expect a crossover between these two regimes to be a generic
feature in critical theories. This suggests that in such gapless
models particular care is required when studying $S_n(\ell)$ for large
$n$.   
\begin{figure}[ht]
\begin{center}
\includegraphics[width=0.7\textwidth]{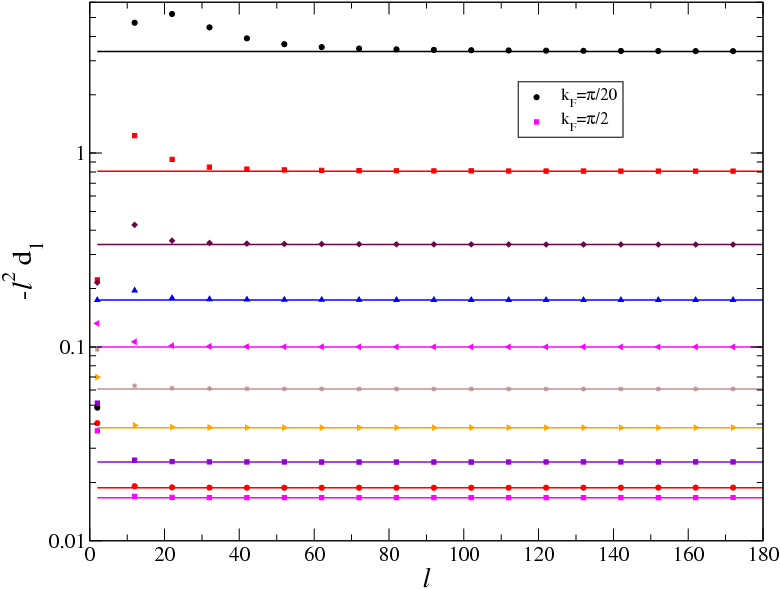}
\end{center}\caption{ 
Correction to scaling for the von Neumann entropy $d_1=S_1-S_1^{JK}$.  
We plot the quantity $-\ell^2 d_1(\ell)$ for values $k_F=n\pi/20$,
where $n=1$ (top curve) up to $n=10$ (bottom curve). The straight
lines are our prediction Eq. (\ref{vNintro}).} 
\label{fig:vN}
\end{figure}
\subsection{Corrections to the von Neumann entropy $S_1(\ell)$}
For the von Neumann entropy $S_1$ all the oscillating terms
vanish and the predicted large-$\ell$ asymptotic behaviour is given in
Eq. (\ref{vNintro}). In Fig.\ref{fig:vN} we plot $-\ell^2 d_1(\ell)$
as a function of $\ell$ and compare it with the prediction
(\ref{vNintro}). We see that the agreement is excellent, which
indicates that further corrections are very small. We note that
for vanishing magnetic field ($k_F=\pi/2$) the amplitude of the
${\cal O}(\ell^{-2})$ correction term is numerically small ($1/60$) so that
the corresponding contribution to $S_1(\ell)$ becomes negligible
already for relatively small $\ell$. At least in the particular case
of the XX model in zero field this shows that the central charge is
most conveniently extracted from finite-size scaling studies of
$S_1(\ell)$ rather than higher R\'enyi entropies.

\subsection{Subleading contributions to $d_n(\ell)$}
\begin{figure}[ht]
\begin{center}
\includegraphics[width=0.7\textwidth]{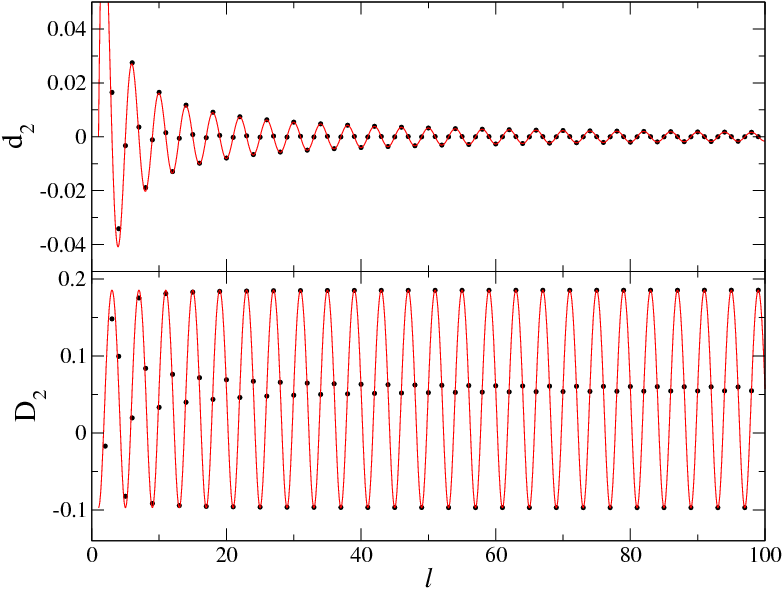}
\end{center}
\caption{Corrections to scaling for the R\`enyi entropy $S_2(\ell)$ 
at $k_F=\pi/4$.\\
\underline{Upper panel}: $d_2(\ell)$ as a function of $\ell$. Dots are the
numerical results while the continuous line corresponds to the
asymptotic expression (\ref{S2}). 
\underline{Lower panel}: rescaled subleading corrections $D_2(\ell)$
defined in (\ref{Dn}) as a function of $\ell$. The agreement between
the asymptotic expression (continuous line) and numerical data (dots)
confirms that (\ref{S2}) is correct to order $o(\ell^{-2})$.} 
\label{fig:Rnc}
\end{figure}

For $n>1$ the structure subleading corrections to scaling for
$S_n(\ell)$ is significantly richer. Explicit expressions, accurate to
order $O(\ell^{-3})$, for the cases $n=2$ and $n=3$ are given in Eqns
(\ref{S2}) and (\ref{S3}) respectively. A comparison of these results
for $n=2$ to numerical computations is presented in
Fig.\ref{fig:Rnc}. We have chosen $k_F=\pi/4$ so as to be able to
separate the oscillation frequencies of the various contributions.
The top panel in Fig. \ref{fig:Rnc} presents a comparison of the
asymptotic expression for $d_2(\ell)$ (continuous lines) to numerical
computations (dots) and show good agreement even for small $\ell$. In
order to better assess the accuracy of the asymptotic results we
introduce the rescaled, subtracted quantity 
\be
D_2(\ell)=[d_2(\ell)- d_2^{\rm asy}(\ell)] \ell^{2}\,,
\label{Dn}
\ee
where $d_2^{\rm asy}(\ell)$ represents the leading correction given in
Eq. (\ref{Sncorrintro}). By construction $D_2(\ell)$ should tend to a
sum of oscillatory terms with fixed amplitudes for large $\ell$. 
The four-sublattice oscillatory behaviour of $D_2(\ell)$ predicted by
(\ref{S2}) is clearly visible and as expected we observe excellent
agreement between the numerical and asymptotic results.
For larger values of $n$, the corrections arising from the `analytic'
part of $D_\ell(\lambda)$ are less important than the harmonic
contributions. Thus the most relevant terms in the asymptotic
expansion are those given in Eq. (\ref{main}). In fact, the first
analytic correction has an exponent $-1-2/n$ and at a given $n$
appears only after $[n/2]$ harmonic contributions with exponents
$-2p/n$ with $p=1,2,\dots$. It has already been observed in Ref.
\cite{ccen-10} that for higher values of $n$ the first subleading
order does not suffice do give an accurate description of the R\'enyi
entropies.  

\begin{figure}[ht]
\begin{center}
\includegraphics[width=0.7\textwidth]{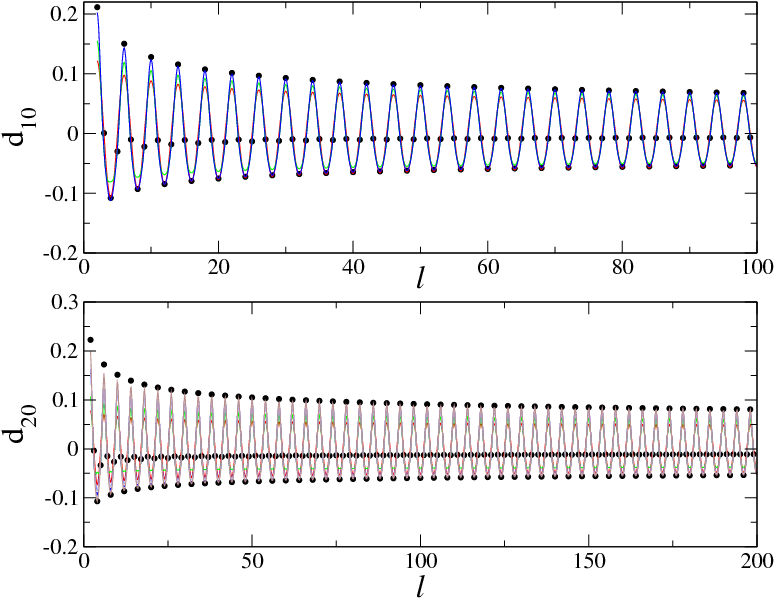}
\end{center}\caption{Corrections $d_n(\ell)$ for $n=10$ (top) and
$n=20$ (bottom) for $k_F=\pi/4$. The numerical data are well
described by Eq. (\ref{main}), but more terms are needed to obtain the
same degree of accuracy when $n$ is increased. 
In both plots the various continuous curves correspond to
Eq. (\ref{main}) one (red curve), two (green curve), three (blue
curve), etc terms in (\ref{main}) retained.
 }
\label{fig:R10-20}
\end{figure}

In Fig. \ref{fig:R10-20} we show a comparison of the corrections
$d_{n}(\ell)$ for $n=10,20$ and $k_F=\pi/4$ with the asymptotic result
Eq. (\ref{main}). Step by step we take into account further terms in
the asymptotic expression (\ref{main}) until we obtain good agreement
with the numerical data. We observe that for $n=10$ three terms in
Eq. (\ref{main}) are enough to reproduces the data, while for $n=20$
we need five terms to have the same accuracy.

\section{Conclusions}
\label{concl}
In this work we have determined the asymptotic behaviour of the
R\'enyi entropies $S_n(\ell)$ in the spin-1/2 XX model for large block
lengths $\ell$. A summary of our results has been presented in
section \ref{sec:summary}. While we have considered the specific case
of the spin-1/2 XX chain in a magnetic field, some features we find
are in fact universal. In particular, the scaling of the leading
oscillatory term (\ref{Sncorrintro}) has been observed for the XXZ
model in zero magnetic field in Ref. \cite{ccen-10}. The corresponding
exponent is modified to $\ell^{-2K/n}$, where $K$ the Luttinger liquid
parameter. This is in full agreement with recent perturbed CFT  
calculations \cite{cc-10}. As we have emphasized repeatedly,
a precise knowledge of the structure of the oscillating terms in
$S_n(\ell)$ is useful for extracting properties such as
the central charge and scaling dimensions of certain operators at
quantum critical points. They furthermore can be used for analyzing
numerical studies of more complicated quantities such as the
entanglement of two disjoint intervals \cite{atc-09,fc-10}.

Oscillating behaviour has also been observed in numerical studies of other
entanglement estimators \cite{ar-10} such as the valence-bond
entanglement. A natural question is whether these can be determined
for certain models using the free fermion techniques we employed for
the XX case as well.

Finally we would like to remark that our results carry over directly
to the critical Ising chain. According to Ref. \cite{ij-08}, the
R\`enyi entropies in the critical Ising model (with $c=1/2$) are
related to those of the spin-1/2 XX chain in zero magnetic field
($k_F=\pi/2$) by 
\be
S_n^{\rm Is}(\ell)=\frac{1}{2}S_n^{XX}(2\ell, k_F=\pi/2).
\ee

\ack
This work was supported in part by the EPSRC under grant
EP/D050952/1 (FHLE) and by the ESF network INSTANS. We thank John
Cardy and Bernard Nienhuis for helpful discussions.
\appendix

\section{Relation between the expansion coefficients for 
$D_\ell$ and  $x_\ell$ }
\label{appA}

In this appendix we report the relations between the coefficients
$a_i$, $b_i$ and $c_i$ in the asymptotic expansion (\ref{DexpT}) of
the Toeplitz determinant $D_\ell(\lambda)$ and the expansion
coefficients $o_{jl}$, $q_{jl}$ characterizing the large-$\ell$
behaviour (\ref{Dexp}) of the auxiliary quantities $x_\ell$.
The following relations hold:
\bea\fl
o_{j1}&=&i(1-2\beta_\l)\cot k_F +\frac{a_1-c_1}2 (2j+1)\,,
\nn\fl
o_{j2}&=&(j+\frac{1}{2})(a_2-c_2)+ (1 - 3 \beta_\l + 3 \beta_\l^2)
+i\cot k_F(2j+1-(j+1)\beta)(a_1-c_1)
\nn\fl&&\quad
+\frac{2j+1}{8}(a_1-c_1)
((2j-1)a_1-(2j+3)c_1)
-\delta_{j0}\left[\frac{1-\beta_\l}{2 \sin k_F}\right]^2\ ,
\nn\fl
q_{10}&=&-\beta_\l^2\ ,\nn\fl
q_{11}&=&\beta_\l^2
\left[\frac{a_1-c_1}2 +i(1-2\beta_\l)\cot k_F\right]-c_1\,,\nn\fl
q_{12}&=&\frac{(a_2-c_2)\beta_\l^2}2-3c_2+\frac{a_1c_1(\beta_\l^2+2)}{4}
-\frac{3a_1^2\beta_\l^2}{8}
+\frac{c_1^2(\beta_\l^2+8)}{8} -  \frac{\beta_\l^2(1 - 2 \beta_\l 
+ 5 \beta_\l^2)}2 \nn\fl&&\quad
+ ((a_1 - c_1) \beta_\l^3 + (1 - 2 \beta_\l) c_1) i \cot k_F 
 + \frac{\beta_\l^2(3 - 14 \beta_\l + 15 \beta_\l^2)}{4\sin^2k_F}\ ,
  \nn\fl
q_{20}&=&0\ ,    \nn\fl
q_{21}&=&\beta_\l^4 \left(\frac{a_1+b_1}2-4\beta_\l i \cot k_F\right)
-\beta_\l^2(\beta_\l^2+1)c_1\,.
\eea
We have derived relations for some other coefficients such as $o_{03}$
and $o_{13}$, but they are not needed for our purposes and we
therefore refrain from reporting them here.

\section{Expansion coefficients for $x_\ell$}
\label{appB}
Substituting the expansion (\ref{Dexp}) into the recurrence
relation (\ref{recx}) gives a set of consistency relations for the
coefficients $q_{jl}$ and $o_{jl}$ in (\ref{Dexp}). These read
\bea
q_{20}&=&q_{30}=q_{40}=q_{21}=q_{31}=q_{41}=0\ ,\\
o_{11}&=&3o_{01}-2(1-2\beta_\l) i \cot k_F\ ,\\
q_{11}&=&q_{10}(-o_{01}+2\beta_\l i \cot k_F)\ ,\\
o_{j1}&=&o_{01}+2j(o_{01}-(1-2\beta_\l) i \cot k_F)\ ,\\
o_{02}&=& \frac{\beta_\l(-1 + 12 \beta_\l - 32 \beta_\l^2 + 27
  \beta_\l^3)}{6}
\nn
&&+\frac{\beta_\l(2 - 13 \beta_\l + 32 \beta_\l^2 - 18 \beta_\l^3)}
{4\sin^2 k_F}\ ,\\
q_{12}&=&\beta_\l^2o_{02}-\beta_\l^4(4+9\beta_\l^2)+\frac{\beta_\l^4}{2\sin^2k_F}(13+18\beta_\l^2)\,.
 \eea

\section{Expressions for the coefficients $B_{pq}$}
\label{app:C}
We recall the expressions for the coefficients $c_{1,2}$ and $b_{1,2}$
(\ref{finalcoef})
\bea
c_1(\beta_\l)&=&2  \beta_\l^3i \cot k_F\,,\nn
c_2(\beta_\l)&=& \frac{\beta_\l^2}6 
(-1 + 7 \beta_\l^2 + 12 \beta_\l^4 - 3 
\beta_\l^2 (5 + 4 \beta_\l^2) \csc^2 k_F) \,,\nn
b_j(\beta_\l)&=&c_j(1+\beta_\l)\ ,\quad j=1,2.
\eea
In terms of the constants $b_{j,q}^+$, $c_{j,q}^+$
\be
b_{j,q}^+=b_j\bigg(\frac{2q-1}{2n}- \frac12\bigg)\ ,\quad
c_{j,q}^+=c_j\bigg(\frac{2q-1}{2n}- \frac12\bigg)\ ,
\quad j=1,2\ .
\ee
the coefficients $B_{pq}$ are given by
\be\fl
B^{(n)}_{p,q}=
2\sin^2(k_F)\left[
b_{2,q}^+-\frac{(b_{1,q}^+)^2}{2}-c_{2,q}^++\frac{(c_{1,q}^+)^2}{2}
+\frac{p}{2}(b_{1,q}^+-c_{1,q}^+)^2\right].
\ee
If $2q-1=n,3n,5n,\ldots$ we instead have $B^{(n)}_{p,q}=0$. An
explicit expression is
\bea
B^{(n)}_{p,q}&=&\frac{x}{3}\left[(5+28x^2)\sin^2(k_F)-15(4x^2+1)\right]\nn
&&
-\frac{p}{4}\left[(1+12x^2)\cos(k_F)\right]^2\Bigg|_{x=\frac{2q-1}{2n}}.
\label{Bpq2}
\eea
\section*{References}

\end{document}